\documentclass[11pt]{article}
\usepackage[T1]{fontenc}
\usepackage{lmodern,graphicx}
\usepackage[margin=1in]{geometry}
\usepackage[hidelinks]{hyperref}
\begin{document}
\begin{center}
{\LARGE\bfseries Who Gets to Do Computing Research? Rethinking Access, Funding, and the Cost of Research\par}
\medskip
{\large How do we ensure that everyone with the ideas, ability, and desire to contribute to computing research can do so?\par}
\medskip
By Gabrielle Allen (University of Wyoming) and Michela Taufer (University of Tennessee Knoxville)\par
Computing Community Consortium (CCC)\quad Nov 13, 2026
\end{center}

The future of computing research will be shaped not just by the questions we ask, but also by who has the opportunity to pursue them. Major investments in advanced computing including artificial/super intelligence (AI) and quantum computing (QC) are expanding opportunities for discovery and innovation. Pursuing these opportunities requires more than a promising idea: researchers also need time, infrastructure, technical expertise, and support to compete for and administer funding.

Access to these resources varies across institutions and may shape who can participate and which research directions they can pursue. A healthy computing research ecosystem depends on research taking place at institutions with different missions and resources, not only at the best-resourced ones.

Computing research spans a broad spectrum of resource requirements, from theoretical work and software developed by small teams to experiments requiring specialized hardware, large datasets, and substantial technical support. Preserving opportunities across this range requires understanding not only the resources needed to conduct research, but also the costs of securing them.

What does it cost an institution — and an individual researcher — to participate in computing research? This article invites computing researchers, university leaders, funding agencies, and industry partners to examine how funding structures, access to infrastructure, administrative responsibilities, and institutional expectations affect research opportunities, and how meaningful pathways can be preserved across institutions with different missions and resources.

\section*{The Changing Path to Funding}

The computing funding landscape is changing. Research spending in computing is growing, but unevenly. According to the \href{https://datavisualization.cra.org/TaulbeeReports/2025/research.html}{2025 CRA Taulbee} \href{https://datavisualization.cra.org/TaulbeeReports/2025/research.html}{Survey}, median research expenditures at Ph.D.-granting computing units rose about 50\% between 2020 and 2025. Over the most recent year, that growth was driven by the best-funded units: the 90th percentile rose from \$28.5 million to \$45.5 million, while the 10th and 25th percentiles stayed nearly flat.

\begin{center}\includegraphics[width=\linewidth]{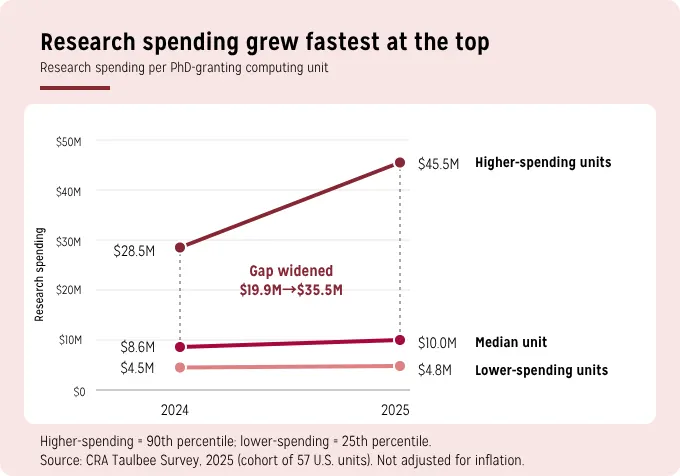}\end{center}

Federal funding shows a parallel \href{https://nsf-gov-resources.nsf.gov/files/FY-2025-Agency-Financial-Report\_0.pdf}{shift}: in FY2025, the National Science Foundation’s (NSF) total funding held steady while its number of new grants fell by about 20\% and its funding rate dropped from 26\% to 19\%. Together, these trends point in the same direction: more of the available funding is flowing to fewer, larger efforts, and to the institutions best equipped to pursue them. Notably, Taulbee covers only Ph.D.-granting units. We have limited systematic data on how these trends affect primarily undergraduate and regional institutions, which both conduct research and introduce many students to it, often students who would not otherwise consider a research career.

Major national investments in areas such as AI and QC can mobilize communities, create infrastructure, support collaborations, and accelerate progress on problems that individual investigators could not address. These investments matter, but their growth also raises questions about the broader ecosystem around them.

Researchers and institutions may increasingly perceive that successful pathways to securing funding depend on alignment with a relatively small number of highly visible national research priorities. Universities may consequently concentrate faculty hiring, infrastructure development, graduate curricula, and internal investment in areas where substantial external funding appears available. This can create \href{https://www.sciencedirect.com/science/article/pii/S0048733325001891?via\%3Dihub}{tension} between national research priorities and the intellectual breadth of computing research.

In this environment, what pathways remain for smaller-scale, exploratory, foundational, unconventional, or emerging research directions? Does a healthy computing research ecosystem require a portfolio of research opportunities supporting different scales and kinds of intellectual contributions?

\section*{Computing Research in EPSCoR States}

This concentration also has a geographic dimension. It is especially evident in NSF EPSCoR (Established Program to Stimulate Competitive Research) states, many of which have substantial rural populations and geographically dispersed higher education systems. Their R1 universities often anchor regional research networks: they collaborate with other universities, community colleges, industry, and government; develop research talent; and contribute expertise to regional industries and communities.

Yet computing research expenditures at these R1 universities are substantially lower than at their non-EPSCoR peers. In our analysis of 146 R1 institutions, pulling from the FY2024 NSF NCSES HERD Survey, the median for EPSCoR institutions was \$6.1M, compared with \$14.9M for non-EPSCoR institutions. Although expenditure differences do not necessarily translate directly into differences in research quality or productivity, this gap may limit their ability to compete for resource-intensive opportunities, recruit and retain faculty and graduate students, and build the infrastructure and partnerships needed to remain competitive.

\begin{center}\includegraphics[width=\linewidth]{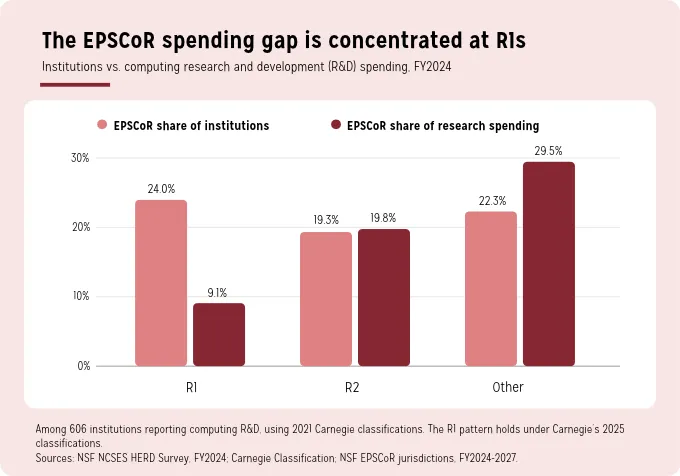}\end{center}

The mechanisms meant to help can be shaped by the same gap. NSF \href{https://www.nsf.gov/science-matters/nsf-101-geographic-diversity-through-epscor}{EPSCoR} co-funding, for example, supports competitive awards in EPSCoR jurisdictions, but leading a large, resource-intensive computing project often requires the very infrastructure and administrative capacity these institutions are still building. Researchers in EPSCoR states may find it easier to join such efforts as partners than to lead them.

The consequences extend well beyond computing departments: As computing, AI, and data-intensive methods become critical across science, engineering, agriculture, healthcare, and other disciplines, universities that fall behind in computing research capacity risk falling behind across their entire research enterprise.

\section*{Access to Computing Infrastructure}

Access to computing resources adds another dimension to the question of who can pursue research opportunities.

Some computing research still has relatively modest resource requirements. Other areas increasingly depend on expensive GPUs, large-scale computing platforms, cloud resources, specialized hardware, very large datasets, high-speed networks, sophisticated software environments, and technical personnel who can operate them. The rapid growth of AI makes this contrast particularly visible, but the issue extends to systems, architecture, networking, cybersecurity, data-intensive science, and other areas of computing.

Some of the most advanced computing capacity, particularly for large-scale AI, now sits in industry. The training compute of notable AI models has been doubling roughly every five months; in 2024, industry \href{https://hai.stanford.edu/ai-index/2025-ai-index-report/research-and-development}{produced} nearly 90\% of notable AI models, and Epoch AI identified none originating from academia. By contrast, a 2024 \href{https://arxiv.org/abs/2410.23261}{survey} of academic AI researchers found that a typical respondent could use about four GPUs at a time, and 85\% reported no cloud-computing budget. Academic researchers have often turned resource constraints into opportunities for creativity, developing ideas that later scale, but industry can now build and operate systems at a scale that few universities can match.

This creates two potentially compounding barriers: researchers may differ both in their ability to secure funding and in their access to the technical infrastructure required to conduct the research itself.

\begin{quote}\itshape Promising research ideas may be constrained not by their intellectual merit but by the institutional environment in which the researcher happens to work.\end{quote}

Shared national resources can help narrow these gaps. NSF’s \href{https://access-ci.org/about/impact/}{Advanced} \href{https://access-ci.org/about/impact/}{Cyberinfrastructure Coordination Ecosystem: Services \& Support (ACCESS)} program offers tiered allocations, from smaller, quickly approved requests suited to a first project to large, competitively reviewed ones, and has served more than 2,000 institutions across all 50 states. The National Artificial Intelligence Research Resource (NAIRR) Pilot \href{https://www.sdsc.edu/news/2026/PR20260901-NAIRR-OC.html}{supported} approximately 900 research teams and educators in all 50 states during its first two and a half years. Yet use remains concentrated: from 2014 to 2023, researchers at R1 universities \href{https://par.nsf.gov/servlets/purl/10579106}{accounted} for about 71\% of users and roughly three-quarters of computing time on these national resources.

These mechanisms are now in transition. The current NSF award supporting ACCESS is scheduled to \href{https://access-ci.org/about/impact/}{end} in 2027, and in September 2026 NSF awarded \$35 million to establish a NAIRR Operations Center, led by the San Diego Supercomputer Center with the Texas Advanced Computing Center, which is now \href{https://par.nsf.gov/servlets/purl/10579106}{defining} how a permanent national AI resource will work. This is the moment to ask whether these mechanisms will reach researchers at institutions with fewer resources, whether those researchers know how to use them, and whether national infrastructure investments will also serve computing research beyond AI.

\section*{The Increasing Cost of Participating in the Research Enterprise}

Conducting research has a cost. But obtaining and administering funding for research also has a cost.

Competing for external funding requires time and institutional capacity before any research begins. Faculty identify opportunities, develop research concepts, build teams, coordinate collaborations, prepare budgets and justifications, develop data and computing plans, navigate institutional requirements, and submit increasingly complex proposals. Some funding opportunities also require institutions to commit cost-share, facilities, or personnel before an award is made, commitments that may be difficult for less-resourced institutions to provide.

Once funding is awarded, additional responsibilities emerge, including financial management, personnel actions, subawards, reporting, compliance, and more. Universities build substantial pre-award and post-award infrastructures to support these activities, but this capacity is not distributed equally across institutions. Indirect cost recovery helps support research administration, but it does not reimburse the effort spent preparing unsuccessful proposals. When institutions lack adequate administrative support, individual faculty and researchers must take on more of this work.

The researcher becomes not only the person developing the intellectual contribution, but also a proposal coordinator, budget planner, project manager, compliance navigator, infrastructure planner, and administrative liaison. For faculty carrying substantial teaching and service responsibilities, this burden can become particularly consequential. The scarce resource, then, is not only money; it is time.

Funding opportunities are also changing. DOE’s Genesis Mission, for example, \href{https://www.energy.gov/articles/energy-department-announces-293-million-funding-support-genesis-mission-national-science}{opened} a \$293 million AI-for-Science funding opportunity in March 2026 with a Phase I deadline about six weeks later and potentially complicated cooperative and project agreements. Fast-turnaround, collaboration-driven opportunities like this can accelerate progress, but may favor institutions with established teams, partnerships, and the administrative capacity to respond quickly.

Programs designed to broaden participation, such as NSF EPSCoR and the CISE Research Initiation Initiative (CRII), recognize that researchers and institutions have different needs and capacities. CRII, for example, has supported early-career faculty at institutions outside the traditional research-intensive university environment. However, CRII’s most recent \href{https://www.nsf.gov/funding/opportunities/crii-computer-information-science-engineering-research-initiation/504952/nsf23-576/solicitation?utm\_source=chatgpt.com}{solicitation} is currently archived and inactive. As support mechanisms tailored to institutions with limited support disappear, the costs described here fall more heavily on the researchers least able to absorb them.

\begin{quote}\itshape This raises an important question for computing: Are we evaluating researchers’ ability to conduct excellent research, or are we increasingly evaluating their — and their institutions’ — capacity to sustain the administrative and infrastructural costs of competing for research funding?\end{quote}

\section*{Research Funding is Not Necessarily Research Success}

These issues are particularly consequential because external research funding plays an important role in faculty evaluation and career advancement. Our own experience suggests that securing grants is often treated as a significant indicator of research excellence, a perception also reported by STEM faculty [Impey et al. (2026)]. At the same time, assistant professors are expected to establish research programs, publish, mentor students, teach significant course loads, fulfill service responsibilities, and continually develop and submit proposals.

When funding is an important measure of research success, researchers need grants to demonstrate their achievements, and competing for those grants requires both individual and institutional investment. This can produce a reinforcing cycle. Institutions with greater research capacity can offer more support to compete for funding, such as proposal staff, cost-share commitments, infrastructure, and time protected from teaching. That support makes successful awards more likely, and those awards bring resources, partnerships, and infrastructure that strengthen the institution’s ability to compete again. Researchers at institutions with fewer resources may have to invest more of their own time simply to enter the same competition.

\begin{center}\includegraphics[width=\linewidth]{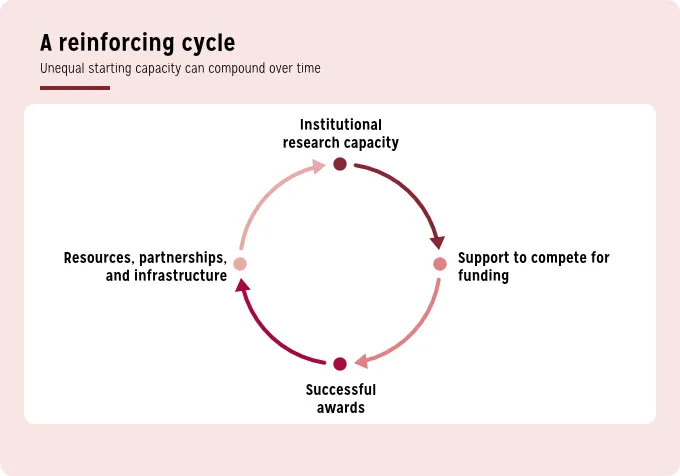}\end{center}

The cycle can also shape what gets studied. When funding is a condition for promotion, faculty may reasonably pursue what is fundable rather than what they consider most important, even though the most significant research is not always the most expensive or the most aligned with current priorities.

Funding totals are, at best, a proxy for what we actually value: research that advances knowledge, trains students for the workforce, and serves the scientific community and society. When a measure becomes a target, it may stop being a good measure. If funding becomes the target, we risk rewarding the capacity to compete for funding rather than the quality and impact of the research itself.

We need to understand whether these dynamics are occurring across the computing research community, how significant they are, and whether their consequences differ across institutions with different missions and resources.

\section*{Questions for the Computing Research Community}

We propose that the computing research community examine several questions.

\textbf{What kinds of computing research do our institutions and funding structures support?}\par Do our definitions and evaluation mechanisms recognize meaningful research conducted at different scales and across different institutional settings?

\textbf{What is the relationship between computing research and research funding?}\par What is funding for? Funding should enable research, not replace it. Do our mechanisms, including those intended to broaden participation in EPSCoR states, help institutions build independent research capacity and leadership, or mainly position them as partners in others’ collaborations?

\textbf{What should research success look like?}

How should we evaluate the quality and impact of research beyond funding totals? When does obtaining funding become a measure of success in itself, and should we expect the same research model from institutions with fundamentally different missions, resources, students, and regional roles?

\textbf{Are major national priorities changing the breadth of computing research?}\par How can we make transformative investments in areas such as AI and QC while preserving opportunities for emerging, foundational, exploratory, and investigator-driven research? And as funding shifts toward fast-turnaround, large-collaboration opportunities, who is positioned to respond?

\textbf{How does access to computing infrastructure affect research opportunity?}\par Who has access to the GPUs, datasets, testbeds, clouds, networks, specialized hardware, software infrastructure, and technical expertise increasingly required by some areas of computing? As national infrastructure transitions and more capacity sits in industry, how can access be extended to researchers at every type of institution?

\textbf{What does it actually cost to participate in computing research?}\par Can we account not only for research expenditures but also for the institutional infrastructure and researcher time required to compete for and administer funding?

\textbf{Who bears those costs?}\par How are pre-award, post-award, administrative, technical, and infrastructure burdens distributed across institutions and individual researchers? Are some researchers discouraged from seeking funding because of the costs of competing?

\textbf{How do teaching responsibilities interact with research expectations?}\par What does it mean to impose similar expectations for externally funded research on faculty operating with very different teaching loads and institutional support?

And perhaps most fundamentally:

\textbf{How do we ensure that everyone with the ideas, ability, and desire to contribute to computing research has a meaningful pathway to do so without requiring every researcher or institution to operate a multimillion-dollar research enterprise?}

\section*{Toward a Community Conversation}

These questions cannot be answered from the perspective of researchers at highly resourced institutions alone. The computing community needs evidence from institutions and researchers operating under different conditions.

Community surveys, interviews, or workshops could test the concerns raised here against researchers’ lived experiences. The answers may confirm or challenge these concerns and reveal institutional models and funding mechanisms that already work. With national infrastructure and funding vehicles in transition, the timing matters.

Before suggesting solutions, we need to understand how funding structures, administrative costs, computing resource requirements, institutional expectations, and faculty workloads affect different types of research. Because in the end, those structures may fundamentally influence who actually gets to do computing research — and whether the future of the computing field will be shaped more by the breadth and quality of its ideas or by the capacity of researchers and their institutions to finance and administer them.

Gabrielle Allen is a Professor of Mathematics and Statistics at the University of Wyoming, where she served as the Founding Director for the School of Computing. She is also adjunct faculty in the Department of Physics and Astronomy. She received a B.S. in Mathematics from Nottingham University in 1988, completed Part III of the Math Tripos (MASt) at Cambridge University in 1989, and received a PhD in Physics from Cardiff University in 1993. She has previously held academic positions in interdisciplinary computational science at Cardiff University, Max Planck Institute for Gravitational Physics, Louisiana State University, Skolkovo Institute of Science and Technology, and the University of Illinois Urbana-Champaign. From 2011 to 2012 she served as a Program Officer at the National Science Foundation in the Office of Cyberinfrastructure. Her research work has focused on the development and application of scientific community software, including the Cactus Framework, Einstein Toolkit, and Grid Application Toolkit. While her work has focused on the modeling of black holes, neutron stars, and gravitational waves, she has worked in diverse disciplinary fields including computer science, education, petroleum engineering, computational chemistry, coastal modeling, and computational fluid dynamics. Dr. Allen was awarded the Gordon Bell Prize in Supercomputing in 2001, the IEEE International Scalable Computing Challenge award in 2009, and the High-Performance Bandwidth Challenge award in 2002. In 2017 she was elected a Fellow of the American Physical Society.

Michela Taufer is an ACM Distinguished Scientist and holds the Jack Dongarra Professorship in High-Performance Computing in the Department of Electrical Engineering and Computer Science at the University of Tennessee, Knoxville (UTK). She earned her undergraduate degrees in computer engineering from the University of Padova (Italy) and her doctoral degree in computer science from the Swiss Federal Institute of Technology or ETH (Switzerland). From 2003 to 2004, she was a La Jolla Interfaces in Science Training Program (LJIS) Postdoctoral Fellow at the University of California, San Diego (UCSD) and The Scripps Research Institute (TSRI), where she worked on interdisciplinary projects in computer systems and computational chemistry. Michela has a long history of interdisciplinary work with scientists. Her research interests include scientific applications on heterogeneous platforms (i.e., multi-core platforms and accelerators); Artificial Intelligence (AI) for cyberinfrastructures (CI); AI integration into scientific workflows, computer simulations, and data analytics. She has been serving as the principal investigator of several NSF collaborative projects. She also has significant experience in mentoring a diverse population of students on interdisciplinary research. Michela’s training expertise includes efforts to spread high-performance computing participation in undergraduate education and research and to increase the interest and participation of diverse populations in interdisciplinary studies.
\end{document}